\documentclass[conference,express_paper]{aesconf} 

\graphicspath{{./}{figures/}}

\usepackage[utf8]{inputenc}

\usepackage{microtype}
\makeatletter
\providecommand\MT@suspend@tagging{}
\providecommand\MT@resume@tagging{}
\makeatother

\usepackage[numbers,square]{natbib}

\usepackage{booktabs}
\usepackage{color}
\usepackage{url}

\usepackage{booktabs, tabularx, enumitem}

\title{An investigation of AI integration in sound designer workflows and experiences.}

\author[1]{Nelly Garcia}
\author[1]{Joshua Reiss}

\affil[1]{Queen Mary University of London}

\correspondence{Nelly Garcia}{n.v.a.garcia-sihuay@qmul.ac.uk}

\lastnames{Garcia and Reiss}

\shorttitle{AI integration in sound designer workflows}

\begin{document}

\twocolumn[
\maketitle 

\begin{onecolabstract}
Artificial intelligence is increasingly being integrated into professional audio production workflows, yet a gap persists between the tools developers produce and the requirements of practising sound designers. This paper investigates this gap through a mixed-methods study comprising a survey of 76 practitioners and follow-up semi-structured interviews with 20 industry professionals. Results were analysed using descriptive statistical analysis and thematic analysis to identify patterns across both datasets.
Five themes emerged from our analysis: Context, Workflow, Potential, Risks, and Right Use. Our work indicates that current AI tools perform adequately in fast-consumption media contexts but lack the narrative sophistication required for high-end sound design (films, immersive experiences etc). Practitioners demonstrate a preference for assistive, task-specific applications, particularly in audio restoration and library management, over end-to-end generative systems. This work contributes to the on-going discussion on the use of AI and AI-enhanced tools in the creative industries. We report on the current status of the field from the point of view of sound designers and creative audio practitioners, and offer a set of recommendation for sound technologist and developers based on our findings to guide the development of more informed AI tools for sound design.

\end{onecolabstract}
]

\section{Introduction}

\begin{figure*}[h]
    \centering
    \includegraphics[width=0.7\linewidth]{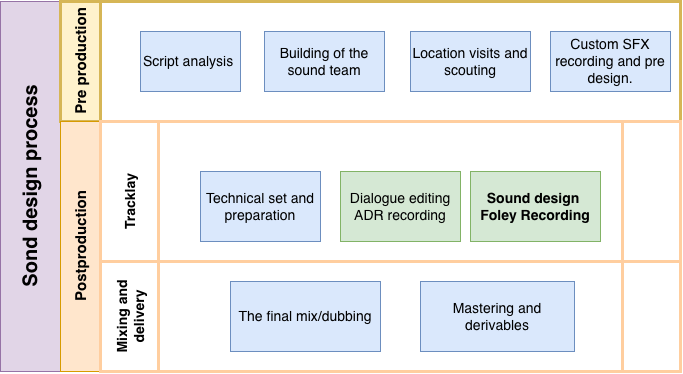}
    \caption{Sound design process, adapted from \cite{sonnenschein2001sound, whittington2007sound}}
    \label{fig:SoundDesignProcess}
\end{figure*}
The history of audio engineering is defined by transformative paradigm shifts, from acoustic to analog recording, from analog to digital, and now from deterministic to generative workflows~\cite{Pavan, Recording, kellogg, morton}. In this third transition, algorithms move beyond discrete tools to operate as active generative agents capable of producing entirely novel sonic material~\cite{pingili, Menexopoulos, oh2023, Camara}.
As AI becomes increasingly embedded across the creative industries, significant uncertainty persists regarding its impact on professional identity~\cite{Katabi, Ewa, hutson} and the skill development of future practitioners~\cite{FREY2017254, Laurie}. In post-production, sound design constitutes the sonic composition of a scene~\cite{Susini, Selfridge}, demanding both rigorous technical knowledge and the aesthetic judgment to serve a narrative moment~\cite{Sara, chion1994audio}. This tension is particularly pronounced in immersive and spatial audio contexts, where parametric precision requirements exceed what current generative models can reliably deliver.
While AI development in audio is accelerating, prevailing research prioritises technical benchmarks and computational efficiency over the practitioner-centred requirements of professional sound design~\cite{MusicLM, liu2023audioldmtexttoaudiogenerationlatent, Sun}. A focus on machine capability alone risks obscuring the craft-based processes through which designers actually work~\cite{Kamath}. The implications of this study are particularly relevant for immersive and game audio contexts. Sound design is not peripheral to immersion, it is constitutive of it. The demands of spatial audio and real-time interaction place distinct requirements on AI tool design that current generative models are not yet meeting.
This paper addresses that gap through a mixed-methods study comprising an online survey of 76 participants and 20 semi-structured interviews, examining how AI is perceived and utilised across diverse professional contexts and the conditions under which these technologies can function as effective collaborative instruments within sound design workflows.
\section{Related Work}

The sound design process begins with a creative partnership between the director and the designer~\cite{whittington2007sound}, progressing from pre-production through asset creation, layering and final mix delivery~\cite{sonnenschein2001sound}, as illustrated in Figure~\ref{fig:SoundDesignProcess}. This study focuses on the technical and creative core of this pipeline (highlighted in green), where designers spend the most time on repetitive, time-consuming tasks, searching for sounds, layering assets, and applying signal processing to serve the scene. In immersive media contexts, spatialisation adds a further layer of technical complexity.
Under accelerating production cycles, generative AI has been positioned as a primary solution to workflow bottlenecks. Viewed through Bruno Latour's Actor-Network Theory~\cite{latour2005reassembling}, these technologies are not merely tools but actants that reshape the creative network. Text-to-audio models enable soundscape construction through natural language prompting~\cite{liu2023audioldm}, while vocal-to-audio tools offer a more performative approach to asset creation~\cite{engel2020ddspdifferentiabledigitalsignal}. However, a significant limitation persists: most generative models carry a music-centric bias, trained on rhythmic and melodic structures~\cite{huang2018music} that poorly translate to the nonlinear, texture-based demands of sound design. As a result, these tools often lack the granular control required by professionals working in complex immersive environments~\cite{fiebrink2018}.
The integration of AI into creative workflows has also raised significant ethical and legal concerns, centred on data provenance and intellectual property~\cite{Herington, Lucchi_2024}. The 'black box' nature of many generative models, trained on copyrighted material without consent or compensation~\cite{sturm2019artificial} — compounds a broader concern regarding the devaluation of professional craft and fair compensation for human expertise.

\section{Methodology}\label{Methodology}

This study employed a mixed-methods research design to capture both the breadth of industry perspectives and the depth of individual professional experience \cite{creswell2017}. The research was divided into two phases: an online survey and online semi-structured interviews.
The survey was distributed via Microsoft Forms\footnote{\url{https://forms.microsoft.com/e/VTprJVvVDC}} over a four-week period, consisting of 15 questions, eight demographic and seven focused on AI integration, workflow optimisation, and ethical positioning.
Semi-structured interviews were conducted as guided conversations, allowing participants to address research topics organically rather than following a fixed sequence of questions \cite{blandford2016qualitative}. The interview guide covered topics including current AI tool usage, perceptions of AI creativity, workflow boundaries, non-generative AI applications, and ethical considerations.

\subsection{Survey}

The survey reached a diverse global audience of 76 participants across 21 countries, with representation spanning the United Kingdom, United States, Mexico, Italy, Germany, Spain and beyond. Gender representation included 51 male, 18 female, 5 non-binary, and 2 who preferred not to say.
The sample was dominated by established professionals: 21 identified as experts, 20 as advanced, and 23 as intermediate, with only 12 identifying as beginners. This is further supported by years of experience, with 38\% possessing over five years in the industry. Participants represented a multi-disciplinary workforce, with sound design as the primary professional identity (60.5\%), alongside significant overlap with music and audio research (39.5\%), post-production (26.4\%), and specialised roles including game audio, Foley and dialogue design.

\subsection{Semi-structured interview}

Twenty practitioners from the survey participated in semi-structured interviews. These were conducted remotely, primarily in English, with the option for Spanish-language sessions to ensure linguistic fluency and precision of technical expression among native speakers. The interviewees were 17 male, 2 female, and 1 non-binary participant, representing 9 sound designers, 4 audio researchers, and 7 music producers across various experience levels. This interdisciplinary composition facilitated a comparative analysis between music production workflows, the narrative-driven requirements of sound design, and the theoretical perspectives of active researchers. All sessions were recorded and transcribed for analysis. Thematic analysis was conducted using NVivo\footnote{\url{https://lumivero.com}} to identify recurring workflow bottlenecks and professional concerns within the socio-technical network \cite{braun2006using}.

 \begin{figure}
    \centering
    \includegraphics[width=0.8\linewidth]{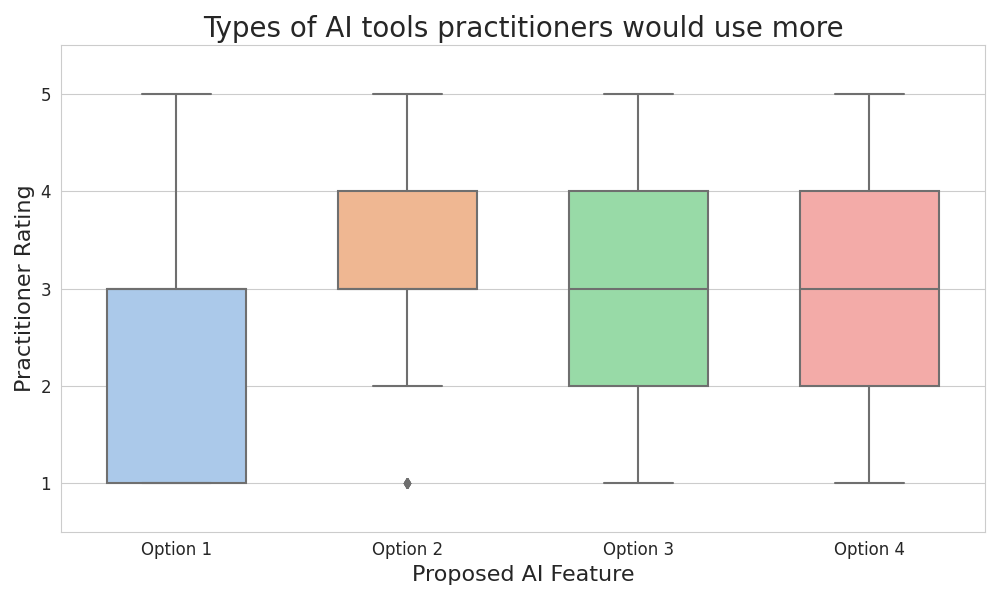}
    \caption{Boxplot representing Likert Scale question in the survey. 5:Most likely to use. 1.Less likely to use. }
    \label{RatingsLikert}
\end{figure}

\section{Quantitative Analysis}
Quantitative analysis of the survey responses presented in Section~\ref{Methodology} reveals a cautious state of transition within the practitioner community. Regarding workflow integration, participants showed a fractured stance: 38\% remained undecided, 34\% expressed willingness to adopt AI tools, and 26\% remained opposed. A notable finding was the high level of technical understanding among respondents. A significant majority (80\%) claimed at least partial understanding of generative and procedural audio processes, suggesting that resistance to AI tools is not born from a lack of understanding but from deliberate professional and ethical positioning. These quantitative patterns informed the themes pursued in the semi-structured interviews, where participants elaborated on the reasoning behind their stances in their own words.
To further characterise tool preferences, participants were asked to rate their likelihood of integrating four proposed AI workflow features (Fig.~\ref{RatingsLikert}): (1) text-to-audio generation via prompting, (2) parametric synthesis engines with real-time control, (3) a hybrid combination of both approaches, and (4) vocal imitation for sound matching and recreation. Practitioners showed a clear preference for tools offering granular control, particularly parametric and hybrid workflows, over fully automated generation. This preference reflects a broader industry demand: sound designers prioritise systems that act as an extension of their technical craft rather than distancing them from the output, suggesting a rejection of full automation in favour of human-in-the-loop co-creation.

A more detailed breakdown of survey responses by region, domain and experience level is reserved for the extended journal version of this work, where the full depth of the dataset will be examined.

\section{Qualitative Analysis}
To further contextualize the survey results, the qualitative phase involved a systematic thematic analysis \cite{guest2012applied,Ronan}of the 20 interview transcripts. This process yielded a hierarchical conceptual framework derived from the participants' shared perspectives. By mapping these professional perspectives, we established five core themes that characterize the intersection of AI and sound design: \textit{Context, Workflow, Risks, Potential, and Right Use}. For a detailed examination, Table \ref{tab:practitioner-themes} provides the definitive framework for these findings, detailing the associated codes and descriptions that underpin each theme.

\begin{table*}[t]
    \centering
    \footnotesize
    \begin{tabularx}{\textwidth}{l >{\raggedright\arraybackslash}p{4.5cm} X}
        \toprule
        \textbf{Theme} & \textbf{Associated Codes} & \textbf{Thematic Description} \\
        \midrule
        \textbf{Context} & 
        \begin{itemize}[leftmargin=*, nosep, after=\vspace{-\baselineskip}]
            \item New reality
            \item Project durability
            \item Post-process
            \item Time-consuming tasks
        \end{itemize} & 
        Explores how sound designers perceive the shifting landscape of their field, specifically identifying which project stages are most influenced by AI integration. \\
        \addlinespace
        \textbf{Workflow} & 
        \begin{itemize}[leftmargin=*, nosep, after=\vspace{-\baselineskip}]
            \item Efficiency \& organization
            \item Source separation
            \item Restoration \& accessibility
            \item Search assistance
        \end{itemize} & 
        Identifies specific pain points and mechanical bottlenecks in current practitioner workflows where AI tools offer potential optimization. \\
        \addlinespace
        \textbf{Risks} & 
        \begin{itemize}[leftmargin=*, nosep, after=\vspace{-\baselineskip}]
            \item Professional unskilling
            \item Fast-consumption media
            \item Lack of trust/adaptability
            \item Agency \& Quality loss
        \end{itemize} & 
        Documents the perceived negative impacts of AI, focusing on the devaluation of craftsmanship and the potential for skill atrophy. \\
        \addlinespace
        \textbf{Potential} & 
        \begin{itemize}[leftmargin=*, nosep, after=\vspace{-\baselineskip}]
            \item Creative inspiration
            \item Decision-making
            \item Iterative drafting
            \item Originality \& Skills
            \item Accessibility
        \end{itemize} & 
        Highlights the perceived positive outcomes of AI as a creative partner that facilitates experimentation and expands the designer's palette. \\
        \addlinespace
        \textbf{Right Use} & 
        \begin{itemize}[leftmargin=*, nosep, after=\vspace{-\baselineskip}]
            \item Global legal frameworks
            \item Copyright \& Ownership
            \item Algorithmic transparency
            \item Permissions \& Regulation
        \end{itemize} & 
        Outlines practitioner requirements for ethical industry standards, emphasizing the need for transparency in training data and legal protection. \\
        \bottomrule
    \end{tabularx}
    \caption{Thematic framework derived from practitioner interviews, mapping the relationship between technical workflows and ethical considerations in AI sound design.}
    \label{tab:practitioner-themes}
\end{table*}
\subsection{Context: Human-in-the-loop}
The Context theme demonstrates that AI adoption in sound design is not a uniform phenomenon, but is contingent on the specific network of production and consumption within which the practitioner operates. The decision to integrate AI therefore functions less as a purely technical choice and more as a negotiation between consumer expectations and the prevailing industry quality standards. The data identifies two distinct production paradigms that condition this negotiation:
\begin{itemize}
\item \textbf{Ephemeral/high-consumption media:} In projects oriented toward rapid turnover and immediate consumption, such as social media or short-form content, efficiency and cost-reduction are the primary drivers. Within these networks, practitioners demonstrate a pragmatic acceptance of turn-key generative tools as a means of bypassing labour-intensive tasks.
\item \textbf{High-Fidelity/long-form production:} Conversely, in projects where long-term cultural durability is the objective, such as cinema or high-end game audio, the human designer retains central mediating authority. Consumer expectations of narrative nuance and emotional specificity demand an artisanal approach that current generative models cannot autonomously replicate.
\end{itemize}

\subsection{Workflow: Optimizing the process}
The production context established in the preceding theme directly conditions the nature of workflow intervention practitioners consider appropriate. Grounded in the traditional sound design pipeline outlined in Section 2, both quantitative and thematic analysis identify specific technical bottlenecks that disproportionately consume time and creative energy across both production paradigms. Practitioners consistently identified a demand for AI intervention in labour-intensive utility tasks, specifically restoration, source separation,  and library management, areas where current generative tooling is perceived as critically underdeveloped relative to professional requirements.

\subsection{Potential Use of AI: Augmentation and Creative Development}
Building on the workflow optimisations identified in the preceding theme, the thematic codes derived from practitioner interviews position the Potential theme as a convergence on the principle of originality through transformation. Rather than supplanting the designer's creative role, AI functions as a means to manipulate parameters, layer complex textures, and traverse libraries at speed, augmenting the iterative process while the practitioner retains final curatorial authority over the output.

Beyond the established practitioner, the data identifies accessibility as a further dimension of this potential. Cloud-based generative tooling materially lowers the barrier to entry for practitioners from contexts where costly studio infrastructure and institutional training have historically been unavailable, extending professional participation to a broader global community. The value of AI-generated material is nonetheless contingent on its passage through the designer's decision-making process, without which the output remains a generic algorithmic product rather than a narratively intentional artefact.

\subsection{Risks: The Artisanal versus The Automated}
The augmentative potential outlined above is, however, conditioned by a significant countervailing risk. The Risks theme and the Potential theme are in direct tension: the same tools that expand creative possibility simultaneously threaten the foundational skill development upon which that creative capacity depends.
The primary concern identified across interview transcripts is professional unskilling(a process whereby sustained reliance on generative tools progressively atrophies the technical abilities that practitioners develop through practice). As AI increasingly automates tasks of moderate complexity, the entry-level work that has traditionally constituted the learning substrate of the discipline is at risk of disappearing. Several participants expressed concern that sound design may consequently retreat into a niche artisanal practice, accessible only to those who acquired foundational skills prior to widespread AI adoption.

\subsection{Right Use: Transparency and data sovereignty}

The tensions identified across the preceding four themes converge in the Right Use theme, which addresses the conditions under which AI integration can be considered ethically legitimate. The data indicates that this legitimacy is not determined by technical capability alone, but by the transparency and provenance of the systems practitioners are asked to adopt.

A primary concern identified across interview transcripts is the black box problem: a widespread lack of clarity regarding the datasets used to train commercial generative models. Compounding this is the issue of data sovereignty, the concern that professional outputs are being harvested to train systems that may subsequently compete for practitioners' contracts. In the absence of opt-in disclosure mechanisms or verifiable copyright clearance for training data, reluctance to fully integrate these tools is positioned by participants not as technophobia, but as a rational professional and ethical response. As one participant observed, engagement with current AI tooling is tempered by a pervasive absence of legal certainty.

Right Use is therefore defined not by what the technology is capable of producing, but by the ethical legitimacy of its development and deployment.

\section{Discussion}
The five themes identified through thematic analysis (\textit{Context, Workflow, Potential, Risks,} and \textit{Right Use}) do not operate independently. The hierarchical structure of the framework (Table~\ref{tab:practitioner-themes}) reveals a progressive chain of interdependency, in which each theme conditions the terms of the next. The central finding is that practitioner resistance to AI is not directed at the technology per se, but at the lack of creative agency within an increasingly automated production pipeline.

\subsection{Context conditions Workflow}

Interpreted through Actor-Network Theory, the two 
production paradigms identified in the Context theme represent fundamentally different network configurations, and these configurations directly determine which workflow interventions practitioners consider appropriate. In ephemeral and high-consumption contexts, AI is adopted primarily to accelerate output and reduce costs, shifting the designer's role from hands-on creator to curator of algorithmically generated material. In high-fidelity contexts, by contrast, the practitioner retains central mediating authority precisely because consumer 
expectations of narrative nuance exceed the autonomous synthesis capacity of current generative models.

This distinction has direct implications for where AI intervention is viable. The Workflow theme data confirms that practitioners do not reject AI categorically, they reject its application in contexts where it cannot meet the cultural and historical specificity the work demands. As illustrated in Figure~\ref{fig:SoundDesignProcess}, the pipeline identifies discrete stages where AI tooling could intervene, yet current generative models remain critically underdeveloped in the areas practitioners need most.

Interviewees mentioned examples of a 1920s Mexican sound palette and a Ugandan market soundscape illustrate this limitation concretely. Generative models trained on generalised datasets cannot autonomously reproduce the granular contextual knowledge that high-fidelity production requires. Workflow optimisation is therefore not a universal objective, but is contingent on production context, either as a means of compressing repetitive, time-consuming procedural tasks, or as a culturally informed creative process that current AI cannot yet support autonomously.

\subsection{Workflow enables Potential, but Potential introduces Risk}

The Workflow theme centres on actions that could reduce time-consuming tasks in the sound design pipeline, specifically utility tasks such as restoration, source separation, and library management. Quantitative survey data corroborates this: Figure~\ref{RatingsLikert} indicates that practitioners rated parametric control and hybrid workflow tools significantly higher than text-to-audio generation, suggesting a preference for assistive intervention over full automation. This finding was reinforced qualitatively through the interviews, where participants consistently identified the same utility tasks as primary bottlenecks, and expressed a desire for tools that compress procedural labour without displacing creative decision-making.

Together, the survey and interview data indicate that AI creates the conditions for the Potential theme to operate. By automating repetitive tasks, practitioners reported being able to redirect cognitive and creative resources toward higher-order aesthetic judgement, what participants described as originality through transformation: the use of AI to manipulate parameters, layer complex textures, and navigate libraries at speed while retaining final curatorial authority.
However, the Risks theme establishes that this augmentative potential carries a structural cost. The same automation that expands creative possibility for established practitioners simultaneously threatens the foundational skill development upon which that creative capacity depends, a concern directly evidenced by the workflow bottlenecks identified in Figure~\ref{fig:SoundDesignProcess}, where the technical and creative core represents precisely the stage at which professional expertise is developed through repetitive practice. Critically, this risk applies even where the accessibility dimension of the Potential theme holds: lowering the barrier to entry through cloud-based tooling does not resolve the unskilling problem if the tools that provide access simultaneously bypass the learning process.

\subsection{Right Use as the governing condition}

The Right Use theme does not follow sequentially from the preceding four, it governs them. The ethical legitimacy of AI integration is the condition under which Context, Workflow, Potential, and Risk can be navigated responsibly. The black box problem and data sovereignty concern identified in the Right Use theme are not peripheral issues; they determine whether practitioners can make informed decisions about when and how to integrate these tools at all. In the absence of verifiable training data provenance and transparent opt-in mechanisms, the rational professional response, as the data confirms, is caution rather than adoption.

Resolving the tensions across these five themes is therefore not a technical problem but a design and governance one. Based on the findings, three recommendations are directed at tool developers:

\begin{enumerate}
    \item AI tools should be designed to enhance the learning curve of sound design, not interrupt it. While this concern was predominantly raised by established  practitioners, the underlying pedagogical principle extends beyond professional self-interest: tools that complete tasks without process transparency risk atrophying the technical intuition and skill formation of the next generation of practitioners, regardless of current expertise level. The most effective integration is therefore one that automates genuinely repetitive and time-consuming tasks, such as stem separation, modification and procedural creation, library management, and automated fade handling, freeing the designer to focus on the aesthetic and narrative decisions that constitute the core of the craft.
    
    \item Every commercial AI audio tool should include a verifiable provenance report detailing the origin and copyright status of its training data, enabling practitioners to make informed ethical decisions and ensuring that integration into professional workflows rests on a foundation of legitimacy. This requirement is increasingly supported by emerging regulatory frameworks: the European Union AI Act (effective August 2026) mandates transparency and labelling for AI-generated content \cite{EUAIAct2024}, with a supporting Code of Practice currently under development that specifically addresses marking and provenance standards for generative audio and music tools \cite{EUCodePractice2025}. The professional community's demand for transparency, as evidenced by this study across 76 practitioners from 21 countries, aligns with these legislative developments, yet also highlights that regulatory frameworks such as the EU AI Act represent only a partial solution. Practitioners from regions without equivalent legislative protection (Uganda, Mexico, Syria and Colombia) expressed identical concerns regarding data provenance and copyright, indicating that current regulatory frameworks do not yet fully address the right use of AI.
    
    \item Where a tool harvests user interactions or creative outputs for model refinement, this must be explicitly disclosed through a transparent opt-in mechanism. Data sovereignty must be recognised and protected as a fundamental professional right.
\end{enumerate}

\section{Conclusion}
This study investigated AI integration in professional sound design workflows through a mixed-methods design comprising a survey of 76 practitioners and 20 semi-structured interviews. The five themes derived from the data indicate that AI's effectiveness is contingent on production context, tool transparency, and the preservation of practitioner agency, rather than generative capability alone. The central gap identified is not technical but philosophical: current tools are oriented toward end-to-end automation rather than the assistive, human-in-the-loop integration practitioners consistently prioritise. Future work should examine how the recommendations proposed here translate into concrete tool design, and longitudinal studies tracking AI adoption across experience levels would strengthen the empirical foundation of this emerging field. The value of AI in sound design is ultimately measured not by the output it generates autonomously, but by the degree to which it extends the practitioner's capacity for deliberate, narratively intentional creative decisions.

\footnotesize
\bibliographystyle{jaes}

\bibliography{refs}

\end{document}